\documentstyle[12pt]{article}
\newcommand{\m}{\medbreak}

\newcommand{\no}{\noindent}
\newcommand{\EQ}{\begin{equation}}
\newcommand{\eq}{\end{equation}}
\newcommand{\EQA}{\begin{eqnarray}}
\newcommand{\eqa}{\end{eqnarray}}

\newcommand{\AR}{\renewcommand {\arraystretch}{1.5}
\begin{array}{l}}
\newcommand{\bAR}{\renewcommand {\arraystretch}{2}
\begin{array}{l}}
\newcommand{\ARc}{\renewcommand {\arraystretch}{1.5}
\begin{array}{c}}
\newcommand{\bARc}{\renewcommand {\arraystretch}{2}
\begin{array}{c}}
\newcommand{\ar}{\end{array} \renewcommand {\arraystretch}{1}}

\newcommand{\ALLPV}{\mbox{$A_{LL}^{PV}\ $}}

\newcommand{\r}{\rightarrow}

\newcommand{\Z}{$Z^{\circ}\ $}
\newcommand{\ZP}{$Z'\ $}

\newcommand{\ET}{\mbox{$E_{T}\ $}}

\begin{document}

\begin{titlepage}
\vspace*{2.5cm}
\begin{center}
{\large \bf  SEARCH FOR QUARK COMPOSITENESS WITH\\
POLARIZED BEAMS AT RHIC \\} 
\vspace*{0.8cm}
{\bf J.M. Virey}{$^1$}  \\ \vspace*{0.6cm}
Centre de Physique Th\'eorique$^{\ast}$, C.N.R.S. - Luminy,
Case 907\\
F-13288 Marseille Cedex 9, France\\ \vspace*{0.2cm}
and \\ \vspace*{0.2cm}
Universit\'e de Provence, Marseille, France\\
\vspace*{6.cm}
{\bf Abstract \\}
\end{center}
Around 1999, thanks to the RHIC Spin Collaboration (RSC),
the Relativistic Heavy Ion Collider (RHIC) will
be used as a polarized proton-proton collider.
A new handed interaction between 
quark subconstituents, which could explain the excess of large $E_T$ jet 
found by the CDF collaboration, could be at
the origin of some small parity violating effects in one-jet
inclusive production.
Using spin asymmetries it is possible, at RHIC, to 
disentangle this new effect from the Standard Model prediction due to 
QCD-ElectroWeak interferences.
\vspace*{1.0cm}

\vfill
\begin{flushleft}

------------------------------------\\
$^{\ast}$Unit\'e Propre de Recherche 7061

{$^1$} Moniteur CIES and allocataire MESR \\
email : virey@cpt.univ-mrs.fr

\end{flushleft}
\end{titlepage}

\section{Introduction}
\indent
\m
The idea of compositeness has been introduced in the hope of solving some
problems of the Standard Model (SM). 
The phenomenological approch is to consider a new "contact" interaction
between quark subconstituents,
which is normalized to a certain
compositeness scale $\Lambda$. This is represented by the following effective 
lagrangian [1] :
\EQ\label{Lcontact}
{\cal L}_{qqqq} = \epsilon \, {g^2\over {8 \Lambda^2}} 
\, \bar \Psi \gamma_\mu (1 - \eta \gamma_5) \Psi . \bar \Psi
\gamma^\mu (1 - \eta \gamma_5) \Psi
\eq
\noindent
where $\Psi$ is a quark doublet, $\epsilon$ is a sign and $\eta$ can take the
values $\pm 1$ or 0. $g$ is a new strong coupling constant
normalized usually to $g^2(\Lambda) = 4\pi$.
\m
Working at Fermilab at the Tevatron ${\bar p} - p$ collider with 
$\sqrt s = 1.8$ TeV, the CDF collaboration has found an excess of events
at high \ET
in the inclusive one-jet cross section
[2]. This can be interpreted as a manifestation of compositeness
for a scale of order
 $\Lambda \sim 1.6$ TeV . In fact we will take this value as the actual limit.
\m 
The expression (eq. \ref{Lcontact}) is rather
general. In particular, there is no reason to assume that the new
interaction is a parity conserving (PC) one.  On the contrary, it has been
advocated for some time [1] that parity
violation (PV)
could be present ($\eta \neq 0$). 

On the other hand, it is well-known from deep-inelastic scattering or
$e^+e^-$ experiments, that the measurement of some spin
asymmetries gives a direct way to pin down a PV interaction. 

It is then tempting to propose the search for an effect which is  
absent in strong processes, like the production of jets,
as long as these processes are solely described in the framework of QCD
which is a parity conserving theory. 
\no
An analysis similar to the one presented here can be found in [3].
\m

The RHIC Spin Collaboration (RSC) [4] has recently proposed 
to run the Brookhaven Relativistic Heavy Ion Collider (RHIC) in the $pp$ mode,
with longitudinally (or transversely) polarized beams. The degree of
polarization of the beam will be as high as 70\%, with a high luminosity of
${\cal L} = 2. 10^{32} \, cm^{-2}.s^{-1}$
at a center-of-mass energy up to 500 GeV.
\m
To build a PV asymmetry one single polarized beam is
sufficient. However, our recent experience [3]
taught us that a larger effect can be
obtained by using both polarized colliding beams which are available at
RHIC. One
can define the double helicity PV asymmetry :
\EQ
\label{ALLPVdef}
A_{LL}^{PV} ={d\sigma_{(-)(-)}-d\sigma_{(+)(+)}\over 
d\sigma_{(-)(-)}+d\sigma_{(+)(+)}}
\eq
\noindent
where the signs  $\pm$ refer to the helicities of the colliding protons. 
\m
A whole set of measurements of the large
Standard PC and PV asymmetries [5] should allow to isolate with very good
precision the
polarized distribution functions of the various partons (quarks and gluons) in a
polarized proton (see refs. [6,7]) and, in the
meantime, to perform some polarization tests of the Standard Model
(for reviews on spin physics at future hadronic colliders one can consult refs. [8,9]).

\m
Let us focus now on the production of a single jet and
concentrate on the high $E_T$ region where quark-quark elastic
scattering is the dominant process and where an effect due to compositeness has
some chance to be observed [3,10].
At RHIC it corresponds to the region $60\, < E_T <\, 120$ GeV. 
\m

We have to remark that we search for some small effects, 
since we want to measure an asymmetry which is zero
according to QCD (QCD is PC !), then the
QCD-Electroweak interference terms 
[11] are no more negligeable.
So the calculations described
below take
into account all the  (lowest order) relevant terms : QCD + Electroweak(EW)
+ Contact
Terms (CT) which are to be added coherently.

\section{Parity violating subprocesses for the one-jet inclusive production}
\indent
\m
If one ignores the contribution of the antiquarks, which is marginal in our \ET
range, and restricts to the main channels $q_iq_i \r q_iq_i$
and $q_iq_j \r q_iq_j$ ($i\neq j$) one gets in short : 
\EQ
\label{ALLPVjet}
A_{LL}^{PV}\ .\ d\sigma \simeq \; \sum_{ij} \sum_{\alpha,\beta}
\int
\left(T_{\alpha,\beta}^{--}(i,j) - T_{\alpha,\beta}^{++}(i,j)
\right) 
\biggl[q_i(x_a)\Delta q_j(x_b) + \Delta q_i(x_a)q_j(x_b) 
+ \;(i\leftrightarrow j) \biggl]
\eq
\noindent
where $T_{\alpha,\beta}^{\lambda_1,\lambda_2}(i,j)$ denotes the matrix element
squared with 
$\alpha$ boson and $\beta$ boson exchanges, or with one exchange process
replaced by a
contact interaction, and 
where we have introduced as usual 
the polarized quark distributions :  
$\Delta q_i(x,Q^2)\ =\ q_{i+} - q_{i-}$, 
$\, q_{i\pm}(x,Q^2)$ being the distributions of 
the polarized quark of flavor $i$, either
with helicity parallel (+) or antiparallel (-) to the parent proton
helicity. Concerning the QCD
contribution to $d\sigma$, we take also into account the antiquarks an also 
$q(\bar q)g$ and $gg$ scattering although these subprocesses are not
dominant in the
high \ET region we consider.

 For the scale $Q^2$, we have taken $Q^2
\, =\,
E_T^2$ after having checked that changing this value between $E_T^2/4$ and
$4\, E_T^2$
has a very small influence on our results on \ALLPV.
\bigbreak
\no
$\bullet $ {\bf Standard QCD-Electroweak interference effects}
\m
At RHIC, a very important parameter is the high luminosity: we obtain indeed
an integrated luminosity
$L_1\ = \ \int{\cal L} dt $ = 800 $pb^{-1}$ after a few
months of run. In the following, we will call $L_2$ the luminosity giving
four times
this sample of events.
We have integrated over a
pseudorapidity interval $\Delta {\eta}^{\star}=1$ centered at ${\eta}^{\star}=0$
, and also over an \ET bin of 10 GeV. The error bars of the
figures are obtained in this context.
\m
Concerning the influence of QCD-EW interference terms 
on \ALLPV, we have already noticed in [3] that more than
90\% of the effect comes both from the interference terms $T_{gZ}$ between
the gluon and \Z (identical quarks) and from the terms
$T_{gW}$  between the gluon and W (quarks of different flavors).
\m
We give in Fig.1 the asymmetry \ALLPV in one-jet production at RHIC which
is expected
from purely QCD-EW interference terms. The correct expressions for the 
$T_{\alpha,\beta}$'s can be found in ref.[6].
We have used various sets of polarized distributions, some quite old, like BRST
[8], CN1 [12] or CN2 [13], and some recent ones which
give better
fits to the new polarized DIS data : BS [14], GS.a,b,c [15] and GRV.s,v [16]. 
Since these latter distributions provide the two extreme cases on \ALLPV 
we will keep only these ones in the following.

We see that \ALLPV remains small, at most
4\% at $E_T=100$ GeV but it is measurable with the sensitivity available
at RHIC. Indeed it lies, with $L_1$, at 2$\sigma$ above zero with GRV or GS
distributions and 4$\sigma$ with BS. 
Finally, the rise of \ALLPV with \ET is due to the increasing importance of quark-quark
scattering relatively to other terms involving gluons. 
\bigbreak
\m
\no $\bullet $ {\bf Interference between Contact and Standard amplitudes}
\m
For the calculation we take all the tree-level diagramms with
gluons, W, \Z , photons exchange and contact term. We have added 
all the terms, involving quarks or antiquarks,
dominant or not. The main
contribution comes from the interference between gluon and contact term,
(due to color rules only identical quark are involved) :

\EQ\label{gCT}
T_{g.CT}^{\lambda_1,\lambda_2}(i,i) \; =\; 
{8 \over 9} \alpha_s\, {\epsilon \over \Lambda^2} \,
(1 - \eta \lambda_1)(1 -  \eta\lambda_2)\left({\hat s^2 \over \hat t} +
{\hat s^2 \over \hat u}\right) 
\eq
\m
Since $\hat t$ and $\hat u$ are negative, $\epsilon = -1\, (+1)$ corresponds
to constructive (destructive) interference [1,3],
and we see that the parameter
governing the sign of \ALLPV is the sign of the product $\epsilon .\eta$ 
(see eq.(\ref{ALLPVjet})).

\section{Discussion and results}
\indent
\m
If we interpret the CDF results as a manifestation of quark substructure 
for a scale $\Lambda = 1.6$ TeV
and want to see the effects of the contact interaction at this scale on
\ALLPV ,
we obtain in Fig. 2 the results of our complete calculation
including all terms. 
The expected Standard asymmetry is shown for comparison.  We have chosen the BS
parametrization for
illustration. One can see that, at RHIC, even with the integrated
luminosity $L_1$,
it is very easy to separate the Standard from the Non-Standard cases, since there 
is a 3$\sigma$ effect with $L_1$.
With GS or GRV distributions, the magnitudes of the asymmetries are reduced but
the effect
is still spectacular.
\m
Now, if we ignore the CDF results and choose
$\Lambda = 2$ TeV (Fig. 3),
we see that, with $L_1$, there is still a 2$\sigma$ difference  
(4$\sigma$ with $L_2$)  between the Standard and Non-Standard asymmetries,
especially for values of \ET above 80 GeV.  
\m
In Fig. 4 we display again \ALLPV with $\Lambda = 2$ TeV, but now
calculated using the
two extreme choices, BS and GS distributions. The clearest result is that,
in spite of
the present uncertainty due to the imperfect knowledge of the polarized quark
distributions, a value for \ALLPV close to zero at large \ET is the sign of
the presence of Non-Standard physics (namely either a left-handed contact
interaction with
destructive interference or a right-handed one with constructive interference).
Indeed, with $L_2$, the two close BS and GS curves stand at $3\sigma$ from
the smaller
QCD-EW asymmetry which corresponds to the GS parametrization. The situation
is less
spectacular in the case where $\epsilon .\eta = -1$ but it is still interesting.
\m

Finally, we have tried to determine if the information from the measurement
of \ALLPV
could compete with the bounds on $\Lambda$ one could reach in the future at the
Tevatron (with unpolarized beams). Following the stategy of refs.
[1,17],  we have calculated $d\sigma$(QCD+CT) in
$p\bar p$
collisions at $\sqrt s = 1.8$ TeV demanding a
100\% deviation from the QCD prediction at large \ET (with at least 10
QCD events). This crude strategy gives amazingly exactly the same result as the
sophisticated CDF study. With an integrated luminosity of 100 $pb^{-1}$ we
expect a limit of  $\Lambda > 2$ TeV at the Tevatron.

Requiering $A_{LL}^{PV} = A_{SM} \pm 2\delta A$ where $\delta A$ is the statistical error,
we obtain at RHIC the following 95\% C.L. limits : with $L_1$ , for any distributions :
$\Lambda \sim 2.0\,$TeV; and with $L_2$ , depending on the distributions :
$\Lambda \sim 2.7 - 3.0\,$TeV.
\m
We see that an increase of the luminosity would increase
considerably the limits on the compositeness scale. Consequently it
appears that the luminosity is a key factor for the polarized analysis.

\section{Conclusion} 
\indent
\m
It has been stressed for some time that polarization at
hadronic colliders should improve their potential capabilities
[8,9], in
particular in the search for New Physics if the energy is as large as the
LHC energy
(see e.g. [18]).
We have seen here that, in spite of its lower energy, the RHIC collider,
running in
the $pp$ mode, could compete with the Tevatron,
thanks to the polarization and also to the high luminosity.
\m
Furthermore, we have shown in a recent paper [19] that a hadrophilic \ZP
[20,21] could induce some visible effects on \ALLPV in the inclusive
one-jet production.
\m
Finally, if the excess of events observed by CDF is due to quark compositeness,
then RHIC will not be a tool for discovery but rather for analysis since it
can provide unique information about the
chirality structure of the new interaction.

\vspace*{1.2cm}
\no {\bf Acknowledgments}

\m

I wish to thank Pierre Taxil for his collaboration.
I am indebted to C. Benchouk,  C. Bourrely, P. Chiappetta, 
M.C. Cousinou, A. Fiandrino, M. Perrottet
and J. Soffer for discussions, help and comments, and to T. Gehrmann and
W.J. Stirling
for providing me some computer program about the GS distributions and to
W. Vogelsang for the GRV one's.
 
\vspace*{0.8cm}
 
\no
[1] E. Eichten, K.Lane and M. Peskin, Phys. Rev. Lett. {\bf 50}, 811 (1983), E.
  Eichten, et al., Rev. Mod. Phys. {\bf 56} (1984) 579.

\no
[2] F. Abe \& al. (CDF Collaboration), FNAL-PUB-96/20-E; A. Goshaw
in these proceedings.

\no
[3] P. Taxil and J.M Virey, Phys. Lett.{\bf B364} 181 (1995) 

\no
[4] RHIC Spin Collaboration (RSC), Letter of intent, April 1991 and RSC
  (STAR/PHENIX) letter of intent update, August 1992.

\no
[5] G. Bunce et al., {\it Polarized protons at RHIC}, Particle World, {\bf 3}, 1
  (1992).

\no
[6] C. Bourrely, J. Ph. Guillet and J. Soffer, Nucl. Phys. {\bf B361}, 72 (1991).

\no
[7] C. Bourrely, and J. Soffer, Nucl. Phys. {\bf B423}, 329 (1994).

\no
[8] C. Bourrely, J. Soffer, F.M. Renard and P. Taxil, Phys. Reports,{\bf 177}, 319
  (1989).

\no
[9] P. Taxil, Riv. Nuovo Cimento, Vol. 16, No. 11 (1993).

\no
[10] M. Tannenbaum, in {\it Polarized Collider Workshop}, J. Collins, S.F.
  Heppelmann and R.W. Robinett eds, AIP Conf. Proceedings {\bf223}, AIP, New
  York, 1990, p. 201.

\no
[11] F.E. Paige, T.L. Trueman and T.N. Tudron, Phys. Rev. {\bf D19}, 935 (1979) ; J.
  Ranft and G. Ranft, Nucl. Phys. {\bf B165}, 395 (1980).

\no
[12] P. Chiappetta and G. Nardulli, Zeit. Phys. {\bf C51}, 435 (1991).

\no
[13] P. Chiappetta, P. Colangelo, J.Ph. Guillet and G. Nardulli, Zeit. Phys. {\bf
  C59}, 629 (1993).

\no
[14] C. Bourrely, and J. Soffer, Nucl. Phys. {\bf B445}, 341 (1995).

\no
[15] T. Gehrmann and W.J. Stirling, Zeit. f. Phys. {\bf C65}, 461 (1995).

\no
[16] M. Gl{\"u}ck, E. Reya, W. Vogelsang, Phys. Lett. {\bf B359}, 201 (1995).

\no
[17] P. Chiappetta and M. Perrottet, Phys. Lett.{\bf B253} 489 (1991) 

\no
[18] A. Fiandrino and P. Taxil, Phys.Rev. {\bf D44}, 3490 (91) and Phys. Lett.
  {\bf B293}, 242 (92).

\no
[19] P. Taxil and J.M Virey, hep-ph 9604331, submitted to Phys. Lett.{\bf B} 

\no
[20] G. Altarelli et al.,
  CERN-TH/96-20 (January 1996) [hep-ph 9601324].

\no
[21] P. Chiappetta et al., PM/96-05,
  CPT-96/P.3304 (January 1996) [hep-ph 9601306];\\ P. Chiappetta in these proceedings.


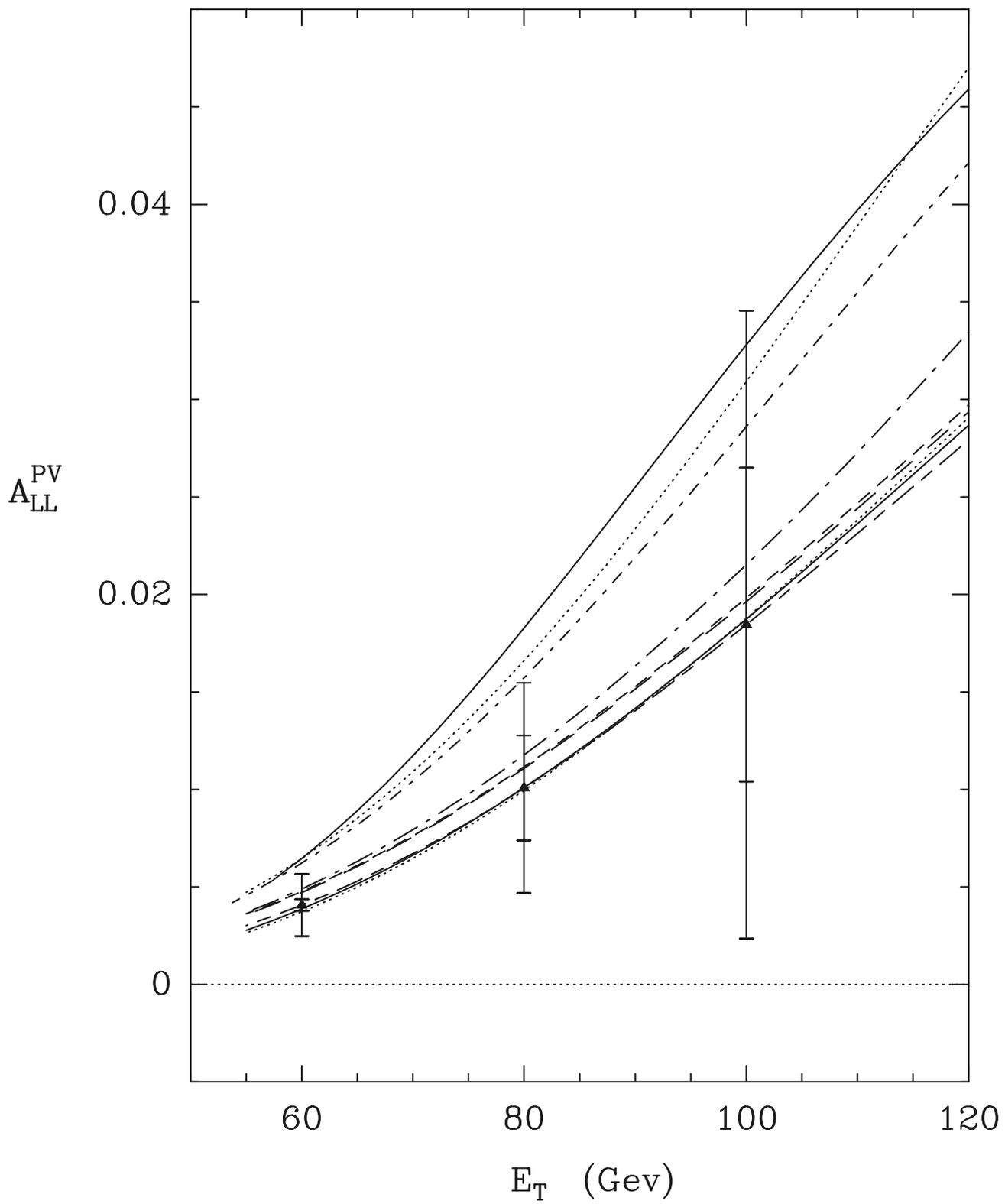
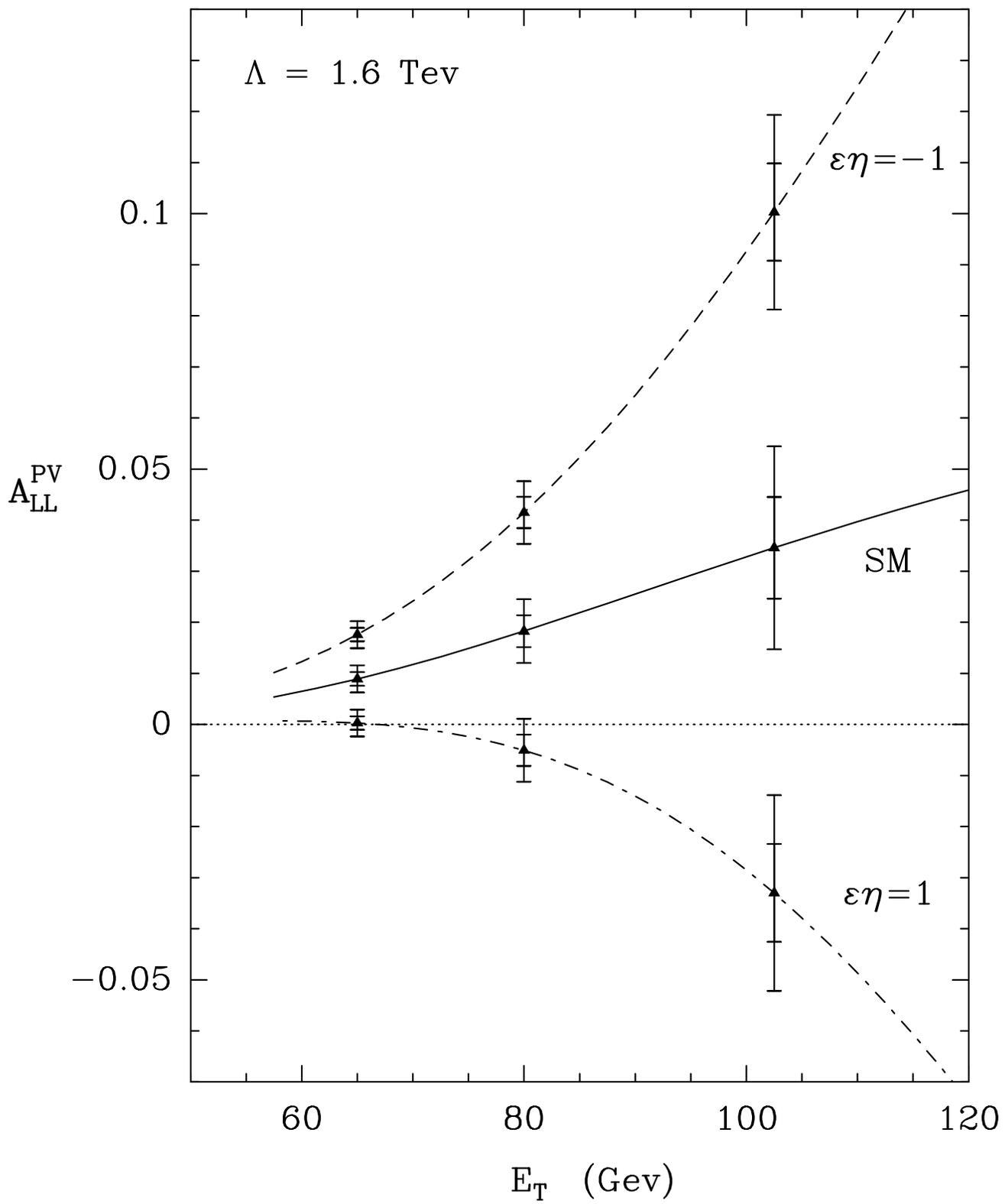
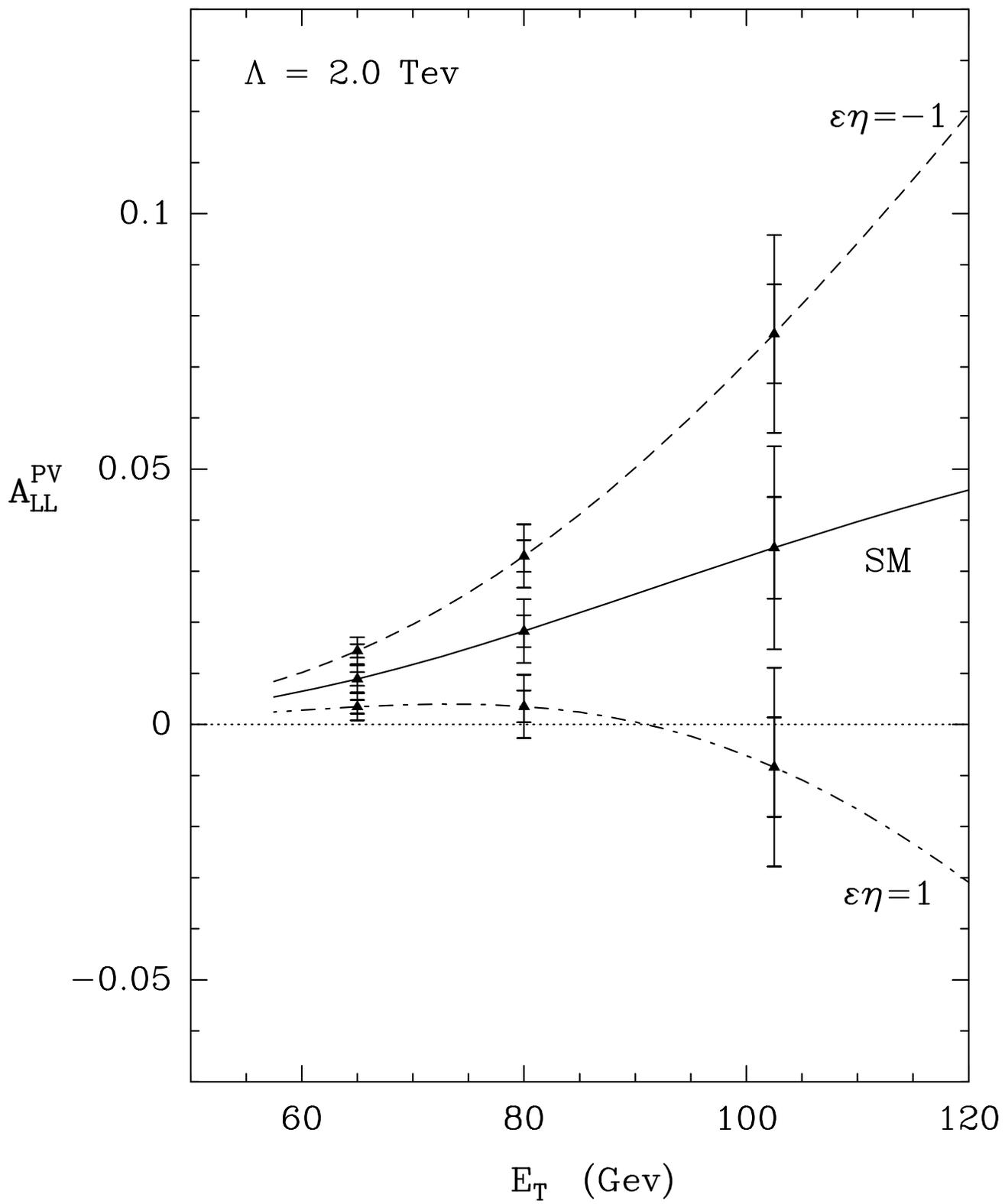
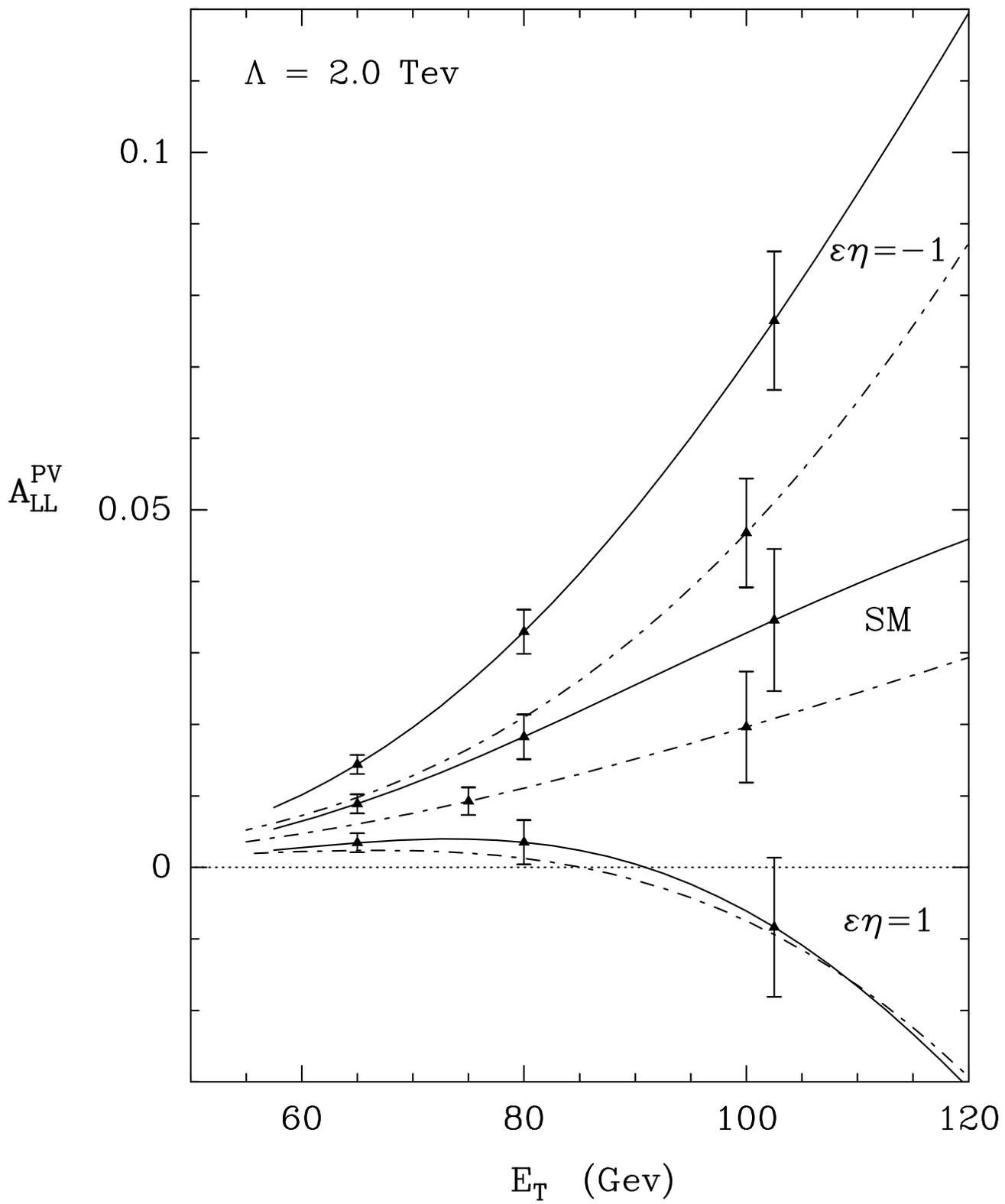
\end{document}